\documentclass[italian,english]{article}
\usepackage[latin9]{inputenc}
\usepackage{color}
\usepackage{amssymb}

\newcommand{\lyxaddress}[1]{
\par {\raggedright #1
\vspace{1.4em}
\noindent\par}
}

\usepackage{babel}

\begin{document}

\title{\textbf{A spherically symmetric and stationary universe from a weak
modification of general relativity}}

\author{\textbf{$^{*}$Christian Corda and $^{+}$Herman J. Mosquera Cuesta}}

\maketitle

\lyxaddress{\begin{center}
\textbf{$^{*}$}Associazione Galileo Galilei, Via Pier Cironi 16
- 59100 PRATO, Italy; \textbf{$^{+}$}Instituto de Cosmologia, Relatividade
e Astrofìsica (ICRA-BR), Centro Brasilero de Pesquisas Fisicas, Rua
Dr. Xavier Sigaud 150, CEP 22290 -180 Urca Rio de Janeiro - RJ Brazil 
\par\end{center}}

\begin{center}
\textit{E-mail addresses:} \textcolor{blue}{christian.corda@ego-gw.it,herman@icra.it} 
\par\end{center}
\begin{abstract}
It is shown that a weak modification of general relativity, in the
linearized approach, renders a spherically symmetric and stationary
model of the universe. This is due to the presence of a third mode
of polarization in the linearized gravity in which a ``curvature''
energy term is present. Such an energy can, in principle, be identified
as the Dark Energy. The model can also help to a better understanding
of the framework of the Einstein-Vlasov system. 
\end{abstract}
PACS numbers 04.40.-b, 04.50.Kd, 04.40.Nr.

The accelerated expansion of the Universe that is currently purported
from observations of SNe Ia suggests that cosmological dynamics is
dominated by a ``new'' substance of the universe constituents dubbed
as Dark Energy, which is able to provide a large negative pressure
to account for the late-time accelerate expansion. This is the standard
picture, in which such a new ingredient is considered as a source
of the \emph{right-hand-side} of the field equations. It is posed
that it should be some form of un-clustered non-zero vacuum energy
which, together with the clustered Dark Matter, drives the global
dynamics. This is the so-called {}``concordance model''($\Lambda$CDM)
which gives, in agreement with the data analysis of the observations
of the Cosmic Microwave Background Radiation (CMBR), Lyman Limit Systems
(LLS) and type la supernovae (SNe Ia), a good framework for understanding
the currently observed Universe. However, the $\Lambda$CDM presents
several shortcomings as the well known``coincidence'' and ``cosmological
constant'' problems \cite{key-1}. 

An alternative approach to explain the purported late-time acceleration
of the universe is to change the \textit{left hand side} of the field
equations, and to inquire whether the observed cosmic dynamics can
be achieved by extending general relativity \cite{key-2,key-3,key-4}.
In this different context, it is not required to search candidates
for Dark Energy and Dark Matter, which until to date, have not been
found, but rather it claims that only the {}``observed'' ingredients:
curvature and baryon matter, have to be taken into account. Considering
this point of view, one can posit that gravity is not scale-invariant
\cite{key-5}. In so doing, one allows for a room for alternative
theories to be opened \cite{key-6,key-7,key-8}. In principle, interesting
Dark Energy and Dark Matter models can be built by considering $f(R)$
theories of gravity \cite{key-5,key-9} (here $R$ is the Ricci curvature
scalar). 

In this perspective, even the sensitive detectors of gravitational
waves like bars and interferometers (i.e., those which are currently
in operation and the ones which are in a phase of planning and proposal
stages\cite{key-10,key-11}, could, in principle, test the physical
consistency of general relativity or of any other theory of gravitation.
This is because in the context of Extended Theories of Gravity important
differences with respect to general relativity show up after studying
the linearized theory\cite{key-12,key-13,key-14,key-15}.

In this letter is shown that a weak modification of general relativity
inspired by f(R) theories of gravity leads to a spherically symmetric
and stationary model of the universe. (The analysis is performed in
the context of the linearized theory). Such a feature appears due
to the presence of a third mode of polarization in the linearized
gravity in which a ``curvature'' energy is present. This peculiar
behavior may in some respect resemble the curvaton field dynamics.
That is, it does not itself drive inflation; it merely generates curvature
perturbations (which could be stationary) at late times after the
inflaton field has decayed and the decay products have redshifted
away, when the curvaton is the dominant component of the energy density.
We recall that the curvaton field is a light scalar field during inflation
whose quantum fluctuations produce the primordial density perturbations
in a scenario for the origin of structure formation. It is posed that
spatial variations in the curvaton density are then transferred to
the cosmic background radiation density when the curvaton decays some
time after inflation. See the following short list of references \cite{key-16}. 

Such an energy can, in principle, be identified as the Dark Energy.
The model can also help to have a better understanding of the Einstein-Vlasov
system \cite{key-17,key-18,key-19,key-20}.

Let us consider the action\begin{equation}
S=\int d^{4}x\sqrt{-g}f_{0}R^{1+\varepsilon}+\mathcal{L}_{m}\label{eq: high order e basta}\end{equation}

Equation (\ref{eq: high order e basta}) is a particular choice in
$f(R)$ theories of gravity \cite{key-2,key-3,key-4,key-5,key-6,key-7,key-9,key-13}. 

In cosmology the action (\ref{eq: high order e basta}) has been analysed
in \cite{key-24} in a rather different cosmological scenario as compared
to the one analysed in this letter. In the limit $\varepsilon\rightarrow0$
and $f_{0}=1$ it recovers the canonical form of the Einstein-Hilbert
action of general relativity \cite{key-21,key-22}, i.e. 

\begin{equation}
S=\int d^{4}x\sqrt{-g}R+\mathcal{L}_{m}\label{eq: standard}\end{equation}

Criticisms on $f(R)$ theories of gravity arises from the fact that
lots of such theories can be excluded by requirements of Cosmology
and Solar System tests \cite{key-23}. However, in the case of the
action (\ref{eq: high order e basta}), the discrepancy with respect
to the standard General Relativity is very weak, because $\varepsilon$
is a very small real parameter. Thus, the mentioned constraints could,
in principle, be satisfied. In particular the authors of \cite{key-23}
found 

\begin{equation}
0\leq\varepsilon\leq7.2*10^{-19}.\label{eq: minimum constrain}\end{equation}
We have also to emphasize that fundamental constrains can be renormalized
in order to obtain $f_{0}=1.$ 

Because we want to study interactions at cosmological scales, the
linearized theory in vacuum, i.e. with $\mathcal{L}_{m}=0$, must
be considered. Notice that it gives a better approximation than the
Newtonian theory \cite{key-28} and the importance of the linearized
theory in a cosmological framework has also been recently emphasized
by George Ellis \cite{key-29}. Therefore, we will analyse the pure
curvature action \begin{equation}
S=\int d^{4}x\sqrt{-g}f_{0}R^{1+\varepsilon}.\label{eq: high order 12}\end{equation}

Also notice that the theory arising from such an action has been recently
linearized in \cite{key-30}, but a review is needed for a better
understanding of the theoretical framework.

By varying the action (\ref{eq: high order 12}) with respect to $g_{\mu\nu},$
the field equations are obtained (through this paper the convention
$G=1$, $c=1$ and $\hbar=1$ will be used) \cite{key-12,key-13,key-30}
\begin{equation}
G_{\mu\nu}=\frac{1}{(1+\varepsilon)f_{0}R^{\varepsilon}}\{-\frac{1}{2}g_{\mu\nu}\varepsilon f_{0}R^{1+\varepsilon}+[(1+\varepsilon)f_{0}R^{\varepsilon}]_{;\mu;\nu}-g_{\mu\nu}\square[(1+\varepsilon)f_{0}R^{\varepsilon}]\}.\label{eq: einstein 2}\end{equation}

By taking the trace of the field equations (\ref{eq: einstein 2})
one gets 

\begin{equation}
\square(1+\varepsilon)f_{0}R^{\varepsilon}=\frac{(1-\varepsilon)}{3}f_{0}R^{1+\varepsilon}.\label{eq: KG}\end{equation}

Then, by making the identifications \cite{key-25,key-30}

\begin{equation}
\begin{array}{ccccc}
\Phi\rightarrow(1+\varepsilon)f_{0}R^{\varepsilon} &  & \textrm{and } &  & \frac{dV}{d\Phi}\rightarrow\frac{(1-\varepsilon)}{3}f_{0}R^{1+\varepsilon}\end{array}\label{eq: identifica}\end{equation}

a Klein - Gordon equation for the effective $\Phi$ scalar field is
obtained. It can be written as

\begin{equation}
\square\Phi=\frac{dV}{d\Phi}.\label{eq: KG2}\end{equation}

To study gravitational waves, the linearized theory has to be analyzed,
with a little perturbation of the background, which is assumed given
by a near Minkowskian background, i.e. a Minkowskian background plus
$\Phi=\Phi_{0}$ (the Ricci scalar is assumed constant in the background)
\cite{key-13,key-25,key-30}. We also assume $\Phi_{0}$ to be a minimum
for the effective potential $V$: 

\begin{equation}
V\simeq\frac{1}{2}\alpha\delta\Phi^{2}\Rightarrow\frac{dV}{d\Phi}\simeq m^{2}\delta\Phi,\label{eq: minimo}\end{equation}

and the constant $m$ has mass dimension. 

Putting

\begin{equation}
\begin{array}{c}
g_{\mu\nu}=\eta_{\mu\nu}+h_{\mu\nu}\\
\\\Phi=\Phi_{0}+\delta\Phi.\end{array}\label{eq: linearizza}\end{equation}

to first order in $h_{\mu\nu}$ and $\delta\Phi$, calling $\widetilde{R}_{\mu\nu\rho\sigma}$
, $\widetilde{R}_{\mu\nu}$ and $\widetilde{R}$ the linearized quantity
which correspond to $R_{\mu\nu\rho\sigma}$ , $R_{\mu\nu}$ and $R$,
the linearized field equations are obtained \cite{key-13,key-25,key-30}:

\begin{equation}
\begin{array}{c}
\widetilde{R}_{\mu\nu}-\frac{\widetilde{R}}{2}\eta_{\mu\nu}=(\partial_{\mu}\partial_{\nu}h_{m}-\eta_{\mu\nu}\square h_{m})\\
\\{}\square h_{m}=m^{2}h_{m},\end{array}\label{eq: linearizzate1}\end{equation}

where 

\begin{equation}
h_{m}\equiv\frac{\delta\Phi}{\Phi_{0}}.\label{eq: definizione}\end{equation}

Then, from the second of eqs. (\ref{eq: linearizzate1}), one can
define the mass like \cite{key-13,key-25,key-30}

\begin{equation}
m\equiv\sqrt{\frac{\square h_{m}}{h_{m}}}=\sqrt{\frac{\square\delta\Phi}{\delta\Phi}}=\sqrt{\frac{\square\delta R^{\varepsilon}}{\delta R^{\varepsilon}}}.\label{eq: massa}\end{equation}

Thus, as the mass is generated by variation of the Ricci scalar, we
can say that, in a certain sense, it is generated by variation of
spacetime curvature, re-obtaining the same result of \cite{key-13,key-25,key-30}. 

Note that, in the present case, the theory is suitable as the modification
of General Relativity is very weak and in agreement with requirements
of Cosmology and Solar System tests \cite{key-23}. 

$\widetilde{R}_{\mu\nu\rho\sigma}$ and eqs. (\ref{eq: linearizzate1})
are invariants under gauge transformations of the type \cite{key-12,key-13,key-30}

\begin{equation}
\begin{array}{c}
h_{\mu\nu}\rightarrow h'_{\mu\nu}=h_{\mu\nu}-\partial_{(\mu}\epsilon_{\nu)}\\
\\h_{\varepsilon}\rightarrow h_{\varepsilon}'=h_{\varepsilon}.\end{array}\label{eq: gauge}\end{equation}

Therefore, one can define \cite{key-30}

\begin{equation}
\bar{h}_{\mu\nu}\equiv h_{\mu\nu}-\frac{h}{2}\eta_{\mu\nu}+\eta_{\mu\nu}h_{\varepsilon}\label{eq: ridefiniz}\end{equation}

and, considering the transform for the parameter $\epsilon^{\mu}$
as

\begin{equation}
\square\epsilon_{\nu}=\partial^{\mu}\bar{h}_{\mu\nu}.\label{eq:lorentziana}\end{equation}
 On this basis one can choose a gauge similar to the Lorentz gauge
that is used when studying electromagnetic waves. It reads: \cite{key-30}

\begin{equation}
\partial^{\mu}\bar{h}_{\mu\nu}=0.\label{eq: cond lorentz}\end{equation}

In this way, the field equations now are given by

\begin{equation}
\square\bar{h}_{\mu\nu}=0\label{eq: onda T}\end{equation}

\begin{equation}
\square h_{\varepsilon}=E^{2}h_{\varepsilon}\label{eq: onda S}\end{equation}

Solutions of eqs. (\ref{eq: onda T}) and (\ref{eq: onda S}) are
plan waves \cite{key-12,key-13,key-30}:

\begin{equation}
\bar{h}_{\mu\nu}=A_{\mu\nu}(\overrightarrow{p})\exp(ip^{\alpha}x_{\alpha})+c.c.\label{eq: sol T}\end{equation}

\begin{equation}
h_{\varepsilon}=a(\overrightarrow{p})\exp(iq^{\alpha}x_{\alpha})+c.c.\label{eq: sol S}\end{equation}

where

\begin{equation}
\begin{array}{ccc}
k^{\alpha}\equiv(\omega,\overrightarrow{p}) &  & \omega=p\equiv|\overrightarrow{p}|\\
\\q^{\alpha}\equiv(\omega_{E},\overrightarrow{p}) &  & \omega_{E}=\sqrt{E^{2}+p^{2}}.\end{array}\label{eq: k e q}\end{equation}

Equation (\ref{eq: onda T}) describes the dynamics of gravitational
waves in standard general relativity \cite{key-21,key-22}. Equation
(\ref{eq: sol T}) gives its solution. In parallel way, equations
(\ref{eq: onda S}) and (\ref{eq: sol S}) describe, respectively,
the dynamics and the solution for the new mode (see also\cite{key-12,key-13,key-30}).

It should be emphasized that the dispersion law for the modes of the
``curvature'' field $h_{\varepsilon}$ is not linear. Besides, the
velocity of every ``ordinary'' wave mode $\bar{h}_{\mu\nu}$ is the
speed of light $c$, i.e., as it arises from general relativity. Rather,
the second equation in (\ref{eq: k e q})) corresponds to the dispersion
law for the mode $h_{\varepsilon}$. Because of this, the mass-energy
field can be described like a wave-packet \cite{key-12,key-13,key-30}.
Also, the group-velocity of a wave-packet of $h_{\varepsilon}$ centered
in in the momentum $\overrightarrow{p}$ is 

\begin{equation}
\overrightarrow{v_{G}}=\frac{\overrightarrow{p}}{\omega},\label{eq: velocita' di gruppo}\end{equation}

which is exactly the velocity of a massive particle with mass-energy
$E$ and momentum $\overrightarrow{p}$.

Therefore, from the second of eqs. (\ref{eq: k e q}) and eq. (\ref{eq: velocita' di gruppo})
it is simple to express the group velocity as:

\begin{equation}
v_{G}=\frac{\sqrt{\omega^{2}-E^{2}}}{\omega}.\label{eq: velocita' di gruppo 2}\end{equation}

As one expects that the wave-packet possesses a constant speed, then
it has to have an energy \cite{key-12,key-13,key-30}

\begin{equation}
E=\sqrt{(1-v_{G}^{2})}\omega.\label{eq: relazione massa-frequenza}\end{equation}

On the other hand, the analysis can remain in the Lorenz gauge with
transformations of the type$\square\epsilon_{\nu}=0$; this gauge
gives a condition of transversal effect for the ordinary part of the
field: $k^{\mu}A_{\mu\nu}=0$, but it does not guarantee the transversal
effect of the total field $h_{\mu\nu}$. From eq. (\ref{eq: ridefiniz})
it becomes

\begin{equation}
h_{\mu\nu}=\bar{h}_{\mu\nu}-\frac{\bar{h}}{2}\eta_{\mu\nu}+\eta_{\mu\nu}h_{\varepsilon}.\label{eq: ridefiniz 2}\end{equation}

At this point, and considering that we are working in the massless
case \cite{key-12,key-13,key-30}, this condition could be expressed
as

\begin{equation}
\begin{array}{c}
\square\epsilon^{\mu}=0\\
\\\partial_{\mu}\epsilon^{\mu}=-\frac{\bar{h}}{2}+h_{\varepsilon},\end{array}\label{eq: gauge2}\end{equation}

which provides the total transversal effect of the field. However,
in the actual (massive) case this is impossible. In fact, by applying
the D'alembertian operator to the second of eqs.(\ref{eq: gauge2})
and using the field equations (\ref{eq: onda T}) and (\ref{eq: onda S})
one arrives to

\begin{equation}
\square\epsilon^{\mu}=E^{2}h_{\varepsilon},\label{eq: contrasto}\end{equation}

which is in contrast with the first of eqs. (\ref{eq: gauge2}). In
the same way it is possible to show that it does not exist any linear
relation between the tensorial field $\bar{h}_{\mu\nu}$ and the {}``curvature''
field $h_{\varepsilon}$. That is why a gauge in which $h_{\mu\nu}$
is purely spatial cannot be chosen (i.e. it cannot be given as $h_{\mu0}=0,$
see eq. (\ref{eq: ridefiniz 2})) . But the traceless condition to
the field $\bar{h}_{\mu\nu}$ can be written as: \cite{key-30}

\begin{equation}
\begin{array}{c}
\square\epsilon^{\mu}=0\\
\\\partial_{\mu}\epsilon^{\mu}=-\frac{\bar{h}}{2}.\end{array}\label{eq: gauge traceless}\end{equation}

These equations imply

\begin{equation}
\partial^{\mu}\bar{h}_{\mu\nu}=0.\label{eq: vincolo}\end{equation}

In order to preserve the conditions $\partial_{\mu}\bar{h}^{\mu\nu}$
and $\bar{h}=0$ one can use transformations of the like

\begin{equation}
\begin{array}{c}
\square\epsilon^{\mu}=0\\
\\\partial_{\mu}\epsilon^{\mu}=0.\end{array}\label{eq: gauge 3}\end{equation}

By taking $\overrightarrow{p}$ in the $z$ direction, a gauge in
which only $A_{11}$, $A_{22}$, and $A_{12}=A_{21}$ are different
to zero can be chosen. The condition $\bar{h}=0$ gives $A_{11}=-A_{22}$.
Now, substituting these equations in eq. (\ref{eq: ridefiniz 2}),
one obtains

\begin{equation}
h_{\mu\nu}(t,z)=A^{+}(t-z)e_{\mu\nu}^{(+)}+A^{\times}(t-z)e_{\mu\nu}^{(\times)}+h_{\varepsilon}(t-v_{G}z)\eta_{\mu\nu}.\label{eq: perturbazione totale}\end{equation}

The term $A^{+}(t-z)e_{\mu\nu}^{(+)}+A^{\times}(t-z)e_{\mu\nu}^{(\times)}$
describes the two standard polarizations of gravitational waves which
arise from General Relativity, while the term $h_{\varepsilon}(t-v_{G}z)\eta_{\mu\nu}$
is the massive field arising from the high order theory. In other
words, the function $R^{\varepsilon}$ of the Ricci scalar generates
a third polarization state for gravitational waves which is not present
in standard general relativity. This third polarization has a {}``curvature''
energy $E.$

Now, a simple model of Universe will be proposed, in which the dynamic
of the matter is described by the Vlasov equation \cite{key-17,key-18,key-19,key-20}.
However, the gravitational forces between the particles, viz., the
galaxies, is now supposed to be mediated by the third mode of eq.
(\ref{eq: perturbazione totale}) after making the assumption that
at cosmological scales such a mode becomes dominant (i.e. $A^{+},A^{-}\ll h_{\varepsilon}$)
\cite{key-28}. In this way the {}``curvature'' energy $E$ can
be identified as the Dark Energy of the Universe $\simeq10^{-29}g/cm^{3}$
\cite{key-28,key-29}. These two assumptions are exacltly the ones
that concern the model of oscillating Universe in \cite{key-28}.

The model that we are going to discuss in this letter is parallel
to the one introduced by Norstrom in \cite{key-19}. All the results
will be obtained adapting the ideas introduced in \cite{key-17,key-18,key-19,key-20}.

In the hypothesis $A^{+},A^{-}\ll h_{\varepsilon}$, the line element
of our model will be the conformally flat one 

\begin{equation}
ds^{2}=[1+h_{\varepsilon}(t,z)](dt^{2}-dz^{2}-dx^{2}-dy^{2}).\label{eq: metrica puramente scalare}\end{equation}

The possible using of a similar conformally flat line element in a
cosmological framework has been discussed in \cite{key-28,key-31}.

To satisfy the condition demanding that the particles make up an ensemble
with no collisions in the spacetime, the particle density must be
a solution of the Vlasov equation

\begin{equation}
\partial_{t}f+\frac{p^{a}}{p^{0}}\partial_{x^{a}}f-\Gamma_{\mu\nu}^{a}\frac{p^{\mu}p^{\nu}}{p^{0}}\partial_{p^{a}}f=0.\label{eq: Vlasov}\end{equation}
Here $\Gamma_{\mu\nu}^{\alpha}$ represent the usual connections,
$f$ is the particle density and $p^{0}$ is determined by $p^{a}$($a=1,2,3$)
according to the relation \begin{equation}
g_{\mu\nu}p^{\mu}p^{\nu}=-1\label{eq: mass-shell}\end{equation}
 \cite{key-17,key-18,key-19,key-20}, which expresses the condition
that the four momentum $p^{\mu}$ lies on the mass shell of the metric
(greek indices run from 0 to 3) \cite{key-17}. 

We recall that, in general, the Vlasov- Poisson system is \cite{key-17,key-18,key-19,key-20}

\begin{equation}
\begin{array}{c}
\partial_{t}f+v\cdot\bigtriangledown_{x}f-\bigtriangledown_{x}U\cdot\bigtriangledown_{v}f=0\\
\\\bigtriangleup U=4\pi\rho\\
\\\rho(t,x)=\int dvf(t,x,v),\end{array}\label{eq: VP}\end{equation}

where $t$ denotes the time and $x$ and $v$ the position and the
velocity of the galaxies. The function $U=U(t,x)$ is the average
Newtonian potential generated by the galaxies. This system represents
the non-relativistic kinetic model for an ensemble of particles with
no collisions, which interacts through the gravitational forces that
they generate collectively \cite{key-17,key-18,key-19,key-20}. Thus,
one can use such a system to describe the motion of galaxies within
the Universe, thought of as pointlike particles, when the relativistic
effects are negligible \cite{key-17,key-18,key-19,key-20}. In this
approach, the function $f(t,x,v)$ in the Vlasov- Poisson system (\ref{eq: VP})
is non-negative and gives the density on phase space of the galaxies
within the Universe.

The Vlasov equation (\ref{eq: Vlasov}) implies that the function
$f$ is constant on the geodesic flow of the line element (\ref{eq: metrica puramente scalare}).
The connection of the line element (\ref{eq: metrica puramente scalare})
are obtained from (note: as we are working in the linearized approach,
in the following computations only terms up to first order in $h_{\varepsilon}$
will be considered while high-order terms will be assumed equal to
zero)

\begin{equation}
\Gamma_{\mu\nu}^{\alpha}=\frac{1}{2}(\delta_{\nu}^{\alpha}\partial_{\mu}h_{\varepsilon}+\delta_{\mu}^{\alpha}\partial_{\nu}h_{\varepsilon}-\frac{1}{1+2h_{\varepsilon}}g_{\mu\nu}\partial_{\alpha}h_{\varepsilon}).\label{eq: connessioni}\end{equation}

In this way, the Vlasov equation in the spacetime defined by the line
element (\ref{eq: metrica puramente scalare}) becomes

\begin{equation}
\partial_{t}f+\frac{p^{a}}{p^{0}}\partial_{x^{a}}f-\frac{1}{2}[2(p^{\mu}\partial_{\mu}h_{\varepsilon})\frac{p^{a}}{p^{0}}+\frac{\partial_{a}h_{\varepsilon}}{(1+2h_{\varepsilon})p^{0}}]\partial_{p^{a}}f=0.\label{eq: Vlasov 2}\end{equation}

Now, let us recall that two quantities are important for the Vlasov
equation in a curve spacetime \cite{key-16}. The first is the current
density

\begin{equation}
N^{\mu}=-\int\frac{dp}{p^{0}}\sqrt{g}p^{\mu}f\label{eq: corrente}\end{equation}

and the second is the stress-energy tensor\begin{equation}
T^{\mu\nu}=-\int\frac{dp}{p^{0}}\sqrt{g}p^{\mu}p^{\nu}f,.\label{eq: energia-impulso}\end{equation}

Here $g$ is the usual determinant of the metric tensor that in the
case of the line element (\ref{eq: metrica puramente scalare}) is
given by

\begin{equation}
g=1+2h_{\varepsilon}.\label{eq: determinante}\end{equation}

Both of $N^{\mu}$ and $T^{\mu\nu}$ are divergence free (conservation
of energy): 

\begin{equation}
\begin{array}{c}
\nabla_{\mu}N^{\mu}\\
\\\nabla_{\mu}T^{\mu\nu}.\end{array}\label{eq: cons. en.}\end{equation}

The mass shell conditions (\ref{eq: mass-shell}) can be rewritten
as 

\begin{equation}
p^{0}=\sqrt{(1+h_{\varepsilon})^{-1}+\delta_{ab}p^{a}p^{b}.}\label{eq: mass-shell 2}\end{equation}

From the connections (\ref{eq: connessioni}), computing the Riemann
tensor, Ricci tensor and Ricci scalar, the {}``effective'' Einstein
field equations \begin{equation}
G_{\mu\nu}=T_{\mu\nu},\label{eq: einstein 3}\end{equation}

can be obtained together with the {}``effective'' Klein - Gordon
equation \begin{equation}
{}\square h_{\varepsilon}=-T,\label{eq: KG 3}\end{equation}

where $T\equiv T_{\mu}^{\mu}$ is the trace of the stress-energy tensor.

To simplify the computations the analysis can be performed in a conformal
frame. Thus, rescaling the stress-energy tensor in the form 

\begin{equation}
T_{*}^{\mu\nu}=(1+3h_{\varepsilon})T^{\mu\nu},\label{eq: riscalo 1}\end{equation}

one obtains\begin{equation}
T_{*}=(1+2h_{\varepsilon})T.\label{eq: riscalo 2}\end{equation}

Note: in general, conformal transformations are performed by rescaling
the line-element like \cite{key-26,key-27}

\begin{equation}
\tilde{g}_{\alpha\beta}=e^{\Phi}g_{\alpha\beta}.\label{eq: conforme}\end{equation}

Here we choose the scalar field as being\begin{equation}
\Phi\equiv h_{\varepsilon},\label{eq: trucco conforme 2}\end{equation}

which also implies \begin{equation}
e^{\Phi}=1+h_{\varepsilon},\label{eq: trucco conforme}\end{equation}

in our linearized approach.

Thus, equation (\ref{eq: KG 3}) becomes

\begin{equation}
{}\square h_{\varepsilon}=-T_{*}.\label{eq: KG 4}\end{equation}

We note that the particle density is still defined on the mass shell
of the starting line element $g_{\alpha\beta}.$ In order to remove
even this last connection with the starting frame we rescale the momentum
as \begin{equation}
p_{*}^{\mu}=(1+\frac{h_{\varepsilon}}{2})p^{\mu}\label{eq: riscalo 3}\end{equation}

and define the particle density in the new conformal frame as

\begin{equation}
f_{*}(t,x,p_{*})=f(t,x,(1-\frac{h_{\varepsilon}}{2})p_{*}).\label{eq: riscalo 4}\end{equation}

Hence, we can write our adaptation of the Vlasov system in the following
form \begin{equation}
{}\square h_{\varepsilon}=(1+2h_{\varepsilon})\int\frac{dp_{*}}{p_{*}^{0}}f_{*}(t,x,p_{*}),\label{eq: KG 5}\end{equation}
\begin{equation}
p_{*}^{0}=\sqrt{1+\delta_{ab}p_{*}^{a}p_{*}^{b}.}\label{eq: mass-shell 3}\end{equation}
\begin{equation}
\partial_{t}f_{*}+\frac{p_{*}^{a}}{p_{*}^{0}}\partial_{x^{a}}f_{*}-\frac{1}{p_{*}^{0}}[p_{*}^{\mu}\partial_{\mu}h_{\varepsilon}p_{*}^{a}+\partial_{a}h_{\varepsilon}]\partial_{p_{*}^{a}}f_{*}=0.\label{eq: Vlasov 3}\end{equation}

Because we want to obtain spherical symmetry to remain in correspondence
with astronomical observations, we can postulate the homogeneity and
isotropy of the Universe. In this case the line-element (\ref{eq: metrica puramente scalare})
assumes the general form \begin{equation}
ds^{2}=[1+h_{\varepsilon}(t,r)](dt^{2}-dr^{2}).\label{eq: metrica puramente scalare 2}\end{equation}

where $r$ is the radial coordinate. Thus, in spherical coordinates,
equations (\ref{eq: KG 5}), (\ref{eq: mass-shell 3}) and (\ref{eq: Vlasov 3})
can be written as \begin{equation}
-\frac{d^{2}h_{\varepsilon}}{dt^{2}}+\frac{1}{r^{2}}\frac{d}{dr}(\frac{d}{dr}h_{\varepsilon}r^{2})=(1+2h_{\varepsilon})\mu(t,r),\label{eq: KG 6}\end{equation}
\begin{equation}
\mu(t,r)=\int\frac{dp}{\sqrt{1+p^{2}}}f(t,r,p),\label{eq: mass-shell 4}\end{equation}
\begin{equation}
\partial_{t}f+\frac{p}{\sqrt{1+p^{2}}}\partial_{r}f-[(\frac{d}{dt}h_{\varepsilon}+\frac{xp}{\sqrt{1+p^{2}}}\frac{1}{r}\frac{d}{dr}h_{\varepsilon})p+\frac{x}{\sqrt{1+p^{2}}}\frac{1}{r}\frac{d}{dr}h_{\varepsilon}]\partial_{p}f=0,\label{eq: Vlasov 4}\end{equation}

where the suffix $*$ has been removed for the sake of simplicity,
and we have denoted by $p$ the vector $p=(p_{1},p_{2},p_{3})$ with
$p^{2}=|p|^{2},$ and also defined $x$ for the vector $x_{i}=(x_{1},x_{2},x_{3})$.

Then, wanting stationary states, following \cite{key-28}, we can
call $\underline{f}$ the frequency of the {}``cosmological'' gravitational
wave (\ref{eq: perturbazione totale}), which has not to be confused
with the particle density that we have previous denoted with $f,$
and assume that $\underline{f}\ll H_{0}$ where $H_{0}$ is the Hubble
constant (i.e. the gravitational wave is {}``frozen'' with respect
the cosmological observations). In this case it is $\frac{d}{dt}h_{\varepsilon}=0$. 

Thus, the system of equations which defines the stationary solutions
of eqs.(\ref{eq: KG 6}), (\ref{eq: mass-shell 4}) and (\ref{eq: Vlasov 4}),
for our model of Universe is \begin{equation}
\frac{1}{r^{2}}\frac{d}{dr}(\frac{d}{dr}h_{\varepsilon}r^{2})=(1+2h_{\varepsilon})\mu(r),\label{eq: KG 7}\end{equation}
\begin{equation}
\mu(r)=\int\frac{dp}{\sqrt{1+p^{2}}}f(r,p),\label{eq: mass-shell 5}\end{equation}
\begin{equation}
p\partial_{r}f-\frac{1}{r}\frac{d}{dr}h_{\varepsilon}[(pr)p+r]\partial_{p}f=0,\label{eq: Vlasov 5}\end{equation}

\subsubsection*{Conclusions}

It has been shown that a weak modification of general relativity,
in the linearized approach produces a spherically symmetric and stationary
model of the universe. This is due to the presence of a third mode
of polarization in the linearized gravity in which a ``curvature''
energy is present. Such an energy can be, in principle, identified
as the Dark Energy. The model can also help to a better understanding
of the framework of the Einstein-Vlasov system.

\subsubsection*{Acknowledgements}

We would like to thank Professor George Ellis, for helpful advices
concerning Cosmology. The Brazilian Section of the International Consortium
of Relativistic Astrophysics, which has financed this research, has
to be thanked too.

Finally, we would like to thank the unknown referee for his enlighten
considerations which permitted to improve this letter.

\end{document}